\documentclass[a4paper,twocolumn,showpacs,superscriptaddress,floatfix]{iopart}

\usepackage{iopams}
\expandafter\let\csname equation*\endcsname\relax
\expandafter\let\csname endequation*\endcsname\relax
\usepackage{amsmath}

\usepackage{soul}

\usepackage{graphicx}
\usepackage{epsf}
\usepackage{epsfig}
\usepackage[usenames]{color}
\usepackage{times}

\usepackage{amssymb}
\usepackage{subfigure}
\usepackage{slashed}
\usepackage{comment}

\newcommand{\be}{\begin{equation}}
\newcommand{\ee}{\end{equation}}
\newcommand{\bea}{\begin{eqnarray}}
\newcommand{\eea}{\end{eqnarray}}
\newcommand{\bal}{\begin{align}}
\newcommand{\eal}{\end{align}}
\newcommand{\nn}{\nonumber}

\font\tenscr=rsfs10 scaled1100
\font\sevenscr=rsfs7 
\font\fivescr=rsfs5 
\skewchar\tenscr='177
\skewchar\sevenscr='177
\skewchar\fivescr='177
\newfam\scrfam
\textfont\scrfam=\tenscr
\scriptfont\scrfam=\sevenscr
\scriptscriptfont\scrfam=\fivescr

\def\scri{{\fam\scrfam I}}

\usepackage{xcolor}
\usepackage{color}

\begin{document}

\title{Pseudospectrum and binary black hole merger transients
}


\author{J.L. Jaramillo}
\address{Institut  de  Math\'ematiques  de  Bourgogne  (IMB),  UMR  5584,  CNRS,\\ Universit\'e  de  Bourgogne  Franche-Comt\'e,  F-21000  Dijon,  France.}

\begin{abstract}
  The merger phase of binary black hole coalescences is a transient between
  an initial oscillating regime (inspiral) and a late exponentially damped phase (ringdown).
  In spite of the non-linear character of Einstein equations, the
  merger  dynamics presents a surprisingly simple behaviour consistent with effective linearity. On the other hand, 
  energy loss through the event horizon and by scattering to infinity renders the
  system non-conservative. Hence, the infinitesimal generator of the
  (effective) linear dynamics is a non-selfadjoint operator.
  Qualitative features of transients in linear dynamics driven by non-selfadjoint (in general, non-normal)
  operators are captured by the pseudospectrum of the
  time generator.
  We propose the pseudospectrum as a unifying framework to thread together the phases of binary black hole
  coalescences, from the inspiral-merger transition up to the late quasinormal mode ringdown.

\end{abstract}



\section{Binary black hole simple dynamics: effective linearity}
\label{s:eff_linear}
Non-linear general relativistic dynamics controlling the binary
black hole (BBH) merger leads to a remarkably simple waveform.
This fact raises the question about
the possibility of describing the dominating features of BBH waveforms in terms of appropriate
effective linear dynamics. Such a perspective has been recently advocated
in \cite{Jaramillo:2022mkh,JarKri22bv2} to address the simplicity and universality of BBH
dynamics. Here we explore such an effective linearity assumption,
when further complemented with the non-conservative character of the
underlying BBH dynamics. Specifically, we propose the non-modal analysis approach
to dynamical transients \cite{TreTreRed93,Schmi07},
built on the notion of pseudospectrum and non-normal linear dynamics~\cite{trefethen2005spectra},
as a systematic and unifying framework for the different regimes of BBH waveforms.

The potential role of effective linearity in BBH dynamics is by no means a new
idea~\footnote{I thank C.F. Sopuerta for stressing, many years ago,
  this important point of BBH dynamics.}.
Early educated expectations based on the non-linearity of
the theory suggested complicated patterns in the BBH merger waveform
(cf. e.g. \cite{Schutz:2004uj}, namely the iconic Thorne's Fig. 1).
This situation changed drastically in 2005 with the  numerical BBH breakthroughs~\cite{Pre05}, 
ultimately confirmed a decade later by observations \cite{Abbott:2016blz},
that revealed a simple BBH waveform.
However, aspects of such a simplicity were 
already present in the previous literature, 
as illustrated by the intuitions coming from the analysis of the GW emission from 
point-particle falling into a
BH~\cite{DavRufPre71} or the Damour's ``effacement property''~\cite{1987thyg.book..128D}
hiding the details in the binary dynamics. 
In particular, the latter stands in a line of thinking culminating
in the Effective-One-Body framework~\cite{Buonanno:2000ef,Buonanno:1998gg},
that provides an effective description of the BBH waveform that
maps the full dynamics to ordinary differential
equations structures and whose extraordinary success provides one of
the strongest indications of the underlying simplicity. Of particular
interest in our present setting is the
``close-limit'' approximation introduced by Pullin
and Price \cite{Price:1994pm,Gleiser:1996yc,Khanna:1999mh}, in which
the merger BBH spacetime is shown to be extremely well captured by the
perturbation of a single black hole, therefore providing
an explicit realisation of `effective linearity' in BBH mergers~\footnote{
Further evidence of such `effective linearity' comes from the Lazarus project
  \cite{Baker:2001nu,Baker:2001sf}  combining results of
  numerical relativity and perturbation theory and, more recently,
  from the `correlation approach' to BH dynamics discussed
  in \cite{Jaramillo:2011re,Jaramillo:2011rf,Jaramillo:2011zw,Jaramillo:2012rr,Gupta:2018znn,Prasad:2020xgr,Iozzo:2021vnq,Ashtekar:2021wld,Ashtekar:2021kqj}, where strong-field spacetime dynamics is probed through the
  cross-correlation of  geometric quantities at the BH horizon and null infinity, which
  implicitly assumes some form of effective linearity.}.
  Most importantly, recent additional support to an effective linear regime
  valid not only at late ringdown BBH waveforms but starting as early as
  the merger~\footnote{Regarding even earlier times, the `simplicity' in the transition from the inspiral to the merger is also illustrated by the accuracy of the Post-Newtonian
    description in the merger, beyond its theoretical limits in the inspiral regime \cite{Borhanian:2019kxt}.}
  is advocated by comparing with accurate full BBH simulations
  in \cite{Giesler:2019uxc} and, crucially, with observations \cite{Isi:2019aib,Capano:2021etf}.
  The latter results are not exempt of controversy (specially,
  regarding the detectability of quasinormal  mode (QNM) overtones \cite{Capano:2021etf,Forteza:2021wfq,Cotesta:2022pci,Isi:2022mhy,Sberna:2021eui,Finch:2022ynt}),  but the idea that some kind of effective
  linearity plays a role in BBH dynamics has indeed entered in current research~\cite{Okounkova:2020vwu}.
  In brief, BBH waveforms seem more `linear' than expected, raising
  the question if the transient phenomenon describing the passage from the inspiral
  to the late ringdown, through the merger phase, can be described in
  terms of effective linear dynamics.

  The last statement is a bold one. Indeed, transients are typically
  associated with non-linear and/or time-dependent dynamics. Along this line, recent insightful
  studies on ringdown dynamics~\cite{Sberna:2021eui} stress precisely the importance of non-linearities
  to account for full BH QNM dynamics. However, this is not in contradiction with the possibility
  that some of the (dominating) qualitative mechanisms may be driven by effectively linear dynamics.
  Even more, one can argue~\footnote{We thank Luis Lehner for his formulation of this problem in his
    presentation at the  BIRS 2022 conference, ``At the Interface of Mathematical Relativity and
    Astrophysics'', Banff (Canada), April 24-29 (2022).}
  that if a sufficiently good control of appropriate underlying background BBH
  dynamics is available, then BBH waveform features may be indeed well described in terms of linear dynamics
  over that dynamical background. The key point is then the correct identification
  of the relevant background dynamics. A particular approach to this problem, based
  on notions of integrability theory, is proposed in Ref.~\cite{JarKri22bv2}.
  Here, we rather adopt an agnostic position regarding the choice of dynamical background, aiming
  at introducing a framework to assess to capability of effective linear dynamics
  to account for BBH transients.

  Specifically we build on the known fact~\cite{trefethen2005spectra} that, in parallel
  and complementarily  to non-linear dynamics, purely linear
  mechanisms can indeed be responsible of transient growth behaviour 
  even if the spectrum of the linearized (time-frozen) operator generating time dynamics
  stays in the stable regime of the complex
  plane. The reason can be traced to a  non-intuitive, but basic fact, in linear
  algebra: linear combinations of exponentially damped eigenvectors can present an
  initial transitory growth before
  fully decaying, as long as they are non-orthogonal (cf. e.g. \cite{Schmi07}).
  In other words, if eigenfunctions~\footnote{For simplicity, diagonalisability is assumed in the
  present discussion.} of the time generator
  are  not orthogonal, purely linear transients can occur. Such non-orthogonality
  of eigenfunctions characterizes the operator as non-normal. In contrast with the
  normal case, in particular for unitary operators in conservative dynamics, the decay properties are
  not fully captured by the spectrum, so a modal (eigenvalue) analysis is not appropriate along the
  full time evolution. The late time dynamics is still controlled by the spectrum, but
  initial and intermediate time regimes require information of the full operator
  in a non-modal analysis approach, intimately related to the notion of
  pseudospectrum~\cite{trefethen2005spectra,Davie07,Sjostrand2019}.
  Such a perspective has proved very illuminating in the
  study of hydrodynamic instability theory and the onset of turbulence in fluid
  dynamics~\cite{TreTreRed93,Schmi07,trefethen2005spectra}, thus
  proving that strong transient growths are not necessarily due to
  non-linearities but can be accounted for by linear dynamics if the
  latter are non-normal.

  Here we explore the applicability of such non-modal approach to BBH merger dynamics.
  Indeed, due to asymptotic radiative losses and flows through the horizon, the effective
  `near-zone' BBH dynamics are non-conservative. This loss of unitarity entails
  the non-selfadjoint character of the infinitesimal time generator that, in
  generic situations, will be a non-normal operator. Under the assumption
  of effective linear BBH dynamics we are therefore in the natural setting to apply
  non-modal analysis to the study of transient growth in BBH mergers.

  \section{Transients and pseudospectrum}
  \label{s:Transients_and_pseudospectrum}

\subsection{Non-normal linear dynamics and transient growth}

Let us consider the following linear dynamical system
\bea
\label{e:dyn_eq}
\left\{
\begin{array}{l}
  \partial_t u =  A u  \\
  u(t=0) = u_0
\end{array}
\right.
\eea
with $A$ an operator acting on functions $u$ in an appropriate Hilbert space $H$,
with scalar product $\langle \cdot, \cdot \rangle$ and norm $||\cdot||$.
We write formally\footnote{This notation is not standard if $A$ is an
  unbounded operator. Here we follow however the presentation and notation in Ref.~\cite{trefethen2005spectra}
  and refer the reader to this reference for technical details
  in the functional analysis treatment.} the solution to the evolution problem as 
\bea
u(t) = e^{tA}u_0 \ .
\eea
We are interested in monitoring the maximum growth rate of solutions $u(t)$ in time, namely
\bea
\label{e:G(t)}
G(t) = \sup_{u_0\neq0}\frac{||u(t)||}{||u_0||} = \sup_{u_0\neq0}\frac{||e^{tA}u_0||}{||u_0||} = ||e^{tA}|| \  ,
\eea
where in the last equation we make use of the operator norm induced from the vector norm $||\cdot||$.
The late time behaviour of $G(t)$ is always described by the spectrum $\sigma(A)$ of $A$. To fix ideas
and assuming $\sigma(A)$ composed only by eigenvalues~\footnote{This is not true in the BH QNM
  case due to the slow decay rate of the effective potential at infinity, as illustrated explicitly
  in the hyperboloidal slicing approach~\cite{Ansorg:2016ztf,PanossoMacedo:2018hab},
  where the continuous part of the spectrum of the (non-selfadjoint)  infinitesimal dynamical
  time generator correspond to tails.
  In this case, eigenvalues in $\sigma(A)$
  (actually, the QNMs) dominate the signal late time behaviour till they are superseded
  by tails.}, at late times $G(t)$ is controlled by
the slowest decaying eigenvalue $\lambda_0$ of $A$, namely
that one with largest real part, so $||e^{tA}||\sim e^{t\mathrm{Re}(\lambda_0)}$.
Late time stability occurs for $\mathrm{Re}(\lambda_0)<0$. However, if we are interested
in $G(t)$ along the whole evolution,
the spectrum $\sigma(A)$ is not enough. Assuming diagonalisability and writing
$A= P \Lambda P^{-1}$, with $\Lambda$ diagonal given by the eigenvalues of $A$ and $P$
constructed from the corresponding eigenfunctions, we can write
\bea
\label{e:G_P}
G(t) = ||e^{tA}|| = ||P e^{t\Lambda}P^{-1}|| \ .
\eea
If $P$ is unitary, i.e. if the eigenfunctions are orthonormal, then indeed
$||P||=1$ (in the norm associated with $\langle \cdot, \cdot \rangle$) and $G(t)$
is well estimated by the spectrum $\sigma(A)$, namely $G(t)\sim || e^{t\Lambda}||$ and
we can do standard spectral (modal) analysis. In particular,
the evolution of linear combinations
of orthogonal eigenfunctions associated with eigenvalues with
negative real part  is monotonically decreasing.
On the contrary, if eigenfunctions of $A$ are non-orthogonal (i.e. if $A$ is non-unitarily
diagonalisable and therefore non-normal), the spectrum $\sigma(A)$ is not enough to
control $G(t)$ and the full operator $e^{tA}$, in particular eigenfunctions of $A$ in $P$
---cf.  Eq.~(\ref{e:G_P})--- must be considered. Crucially, the evolution of
linear combinations
of non-orthogonal eigenfunctions  can lead to transient growths, in a purely linear mechanism,
even if associated eigenvalues have negative real part: a genuine non-modal analysis
phenomenon.

\subsection{Non-modal analysis: spectrum, numerical range and pseudospectrum}
Three sets in the complex plane $\mathbb{C}$, associated with $A$, play a relevant role in the
discussion:
the spectrum  $\sigma(A)$, the numerical range $W(A)$
and, fundamentally, the $\epsilon$-pseudospectra $\sigma^\epsilon(A)$. Of special relevance are
the suprema of the real parts of each respective set. Specifically:
\begin{itemize}
\item[i)] {\em Spectrum $\sigma(A)$}. Namely the set of $\lambda\in \mathbb{C}$ where
  the resolvent $R_A(\lambda) = (A -\lambda I)^{-1}$ is not defined as a bounded operator.
  The spectral abscissa $\alpha(A)$ is defined by  
  \bea
  \label{e:spectral_abscissa_def}
\alpha(A) &=&\sup \mathrm{Re}\big(\sigma(A)\big)
= \sup\{\mathrm{Re}(\lambda), \hbox{with} \ \lambda\in\sigma(A)\} \ .
\eea
In the QNM context $\alpha(A)$ corresponds to the so-called {\em spectral gap}.

\item[ii)] {\em Numerical range $W(A)$}. This set depends on the scalar product and is defined as
  \bea
  \label{e:W(A)}
W(A) = \{\langle u, A u\rangle, \hbox{with} \ ||u||=1 , u\in H\} \ .
\eea
The numerical abscissa $\omega(A)$ is defined as 
\bea
\label{e:numerical_abscissa_def}
\omega(A) &=&\sup \mathrm{Re}\big(W(A)\big) 
= \sup\{\mathrm{Re}(\lambda), \hbox{with} \ \lambda\in W(A)\} \ .
\eea
A very useful characterization, cf. e.g. \cite{trefethen2005spectra}, is (here $ A^\dagger$
is the adjoint of $A$ through $\langle \cdot, \cdot \rangle$)
  \bea
  \label{e:numerical_abscissa_characterization}
   \omega(A) = \sup \sigma\left(\frac{1}{2}(A + A^\dagger)  \right) \ .
  \eea

\item[iii)] {\em $\epsilon$-Pseudospectrum $\sigma^\epsilon(A)$}. This is the fundamental notion in the present
  setting. As $W(A)$ and in contrast with the spectrum, it depends on the
  choice of scalar product~\cite{Gasperin:2021kfv}. It admits different characterizations,
  each one stressing a complementary aspect
  \bea
\label{e:pseudospectrum_def}
\sigma^\epsilon(A) &=& \{\lambda\in\mathbb{C}:
||R_A(\lambda)|| = ||\lambda\mathrm{Id}- A)^{-1}||>1/\epsilon\}  \\
&=& \{\lambda\in\mathbb{C}, \exists v\in H: ||A v-\lambda
v||<\epsilon, ||v||=1\} \nn \\
&=&\{\lambda\in\mathbb{C}, \exists \; \delta
A \ \hbox{bounded operator}, ||\delta A||<\epsilon:
\lambda\!\in\!\sigma(A+\delta A) \} \nn 
\eea
The first one characterizes $\epsilon$-pseudospectra as sets bounded by the contour levels
of the norm of the resolvent $R_A(\lambda)$ (as a function in the $\lambda$-complex plane),
the second introduces the notion of $\epsilon$-quasimode $v$, whereas the latter permits
to address the spectral instability under perturbations $\delta A$ of $A$.
  The key feature in the
present  setting is that, even if the spectrum $\sigma(A)$ lays in the stable left-half
of $\mathbb{C}$, growth transients can happen if the $\epsilon$-pseudospectra $\sigma^\epsilon(A)$
sets protrude significantly in the unstable right-half plane.
This leads to the introduction, for each $\epsilon>0$, of the pseudospectral abscissa
\bea
\label{e:pseudospectral_abscissa_def}
\alpha_\epsilon(A) &=&\sup \mathrm{Re}\big(\sigma^\epsilon(A)\big)
= \sup\{\mathrm{Re}(\lambda), \hbox{with} \ \lambda\in \sigma^\epsilon(A)\} \ .
\eea
\end{itemize}

\noindent The relevance of these three sets is that they control, at different time regimes,
the maximum  growth rate $G(t)=||e^{tA}||$
of solutions $u(t)$ to Eq. (\ref{e:dyn_eq}). Specifically, the spectrum $\sigma(A)$
controls the late $t\to\infty$ limit through the spectral abscissa
$\alpha(A)$, whereas the  numerical range $W(A)$ controls the initial growth as $t\to 0^+$,
through the numerical abscissa $\omega(A)$. The key structural role of the $\epsilon$-pseudospectrum notion
arises from the fact that it interpolates between late times controlled by $\sigma(A)$ (in the $\epsilon\to 0$
limit) and early times controlled by $W(A)$ (in the $\epsilon\to \infty$ limit),
providing a control of the growth rate at intermediate times in terms of the
pseudospectral abscissa $\alpha_\epsilon(A)$, for the whole range $\epsilon \in \; ]0,\infty[$.
  In this sense, since it takes into account the full structure of $A$ and not only its
spectrum, the $\epsilon$-pseudospectrum is the relevant object to consider in generic non-normal
dynamics, recovering the standard spectral approach for normal (e.g. unitary) dynamics
by taking $\epsilon\to 0$, since $\sigma(A)=\lim_{\epsilon\to 0}\sigma_\epsilon(A)$.

\subsection{The transient dynamical regimes and the pseudospectrum}
\label{s:transient_pseudospectrum}
We make explicit now the relation of $\sigma(A)$, $W(A)$ and $\sigma_\epsilon(A)$
with respective time regimes:

\begin{itemize}

\item[i)] {\em Triggering of the transient: numerical abscissa $\omega(A)$}.
  
  The initial slope in the maximal growth factor $G(t)=||e^{tA}||$ is given by
  \cite{trefethen2005spectra}
  \bea
  \label{e:numerical_abscissa_early_behaviour}
  \frac{d}{dt}\left.||e^{tA}||\right|_{t=0} =
  \lim_{t\to 0^+} \frac{1}{t}\ln ||e^{tA}|| = \omega(A) \ .
  \eea

  As indicated above, this initial growth is controlled by $\epsilon$-pseudospectra
  in the $\epsilon\to\infty$ limit, specifically by the depart 
  of the pseudospectral abscissa from an $\epsilon$-linear behaviour 
  \bea
  \label{e:epsilon_to_infty}
  \omega(A)= \lim_{\epsilon \to \infty} \left(\alpha_\epsilon(A) - \epsilon\right) \ .
  \eea
  This confers the numerical abscissa $\omega(A)$ with the significant role
  of detecting the possibility of the triggering of a growth transient.
  Moreover, using the characterization  of $\omega(A)$
  in (\ref{e:numerical_abscissa_characterization}), in the case that $\omega(A)$
  is realised as a maximum and not only the supremum, the corresponding
  eigenvector $u_{\omega(A)}$ of the maximum eigenvalue --- i.e. $\omega(A)$ --- in 
  \bea
  \left(\frac{1}{2}(A + A^\dagger\right)u_{\omega(A)} = \omega(A)u_{\omega(A)} \ ,
  \eea
  realizes the maximum possible growth in  $||e^{tA}||$ at $t=0$. Thus, the choice of initial data
  \bea
  \label{e:maximum_growth_initialdata}
  u_0 := u_{\omega(A)} \ ,
  \eea
  in Eq. (\ref{e:dyn_eq})
  provides the candidate for the initial data
  with strongest initial growth
  transient.
     
  Finally, the numerical abscissa provides an upper bound for the growth factor
   \cite{trefethen2005spectra}
   \bea
   \label{e:eA_upper_bound}
  ||e^{tA}||\leq e^{t\omega(A)} \ , \ \forall t\geq 0 \ ,
  \eea
  holding along the transient. In particular, $\omega(A) =0$ characterises $A$ as 
  contractive
  \bea
  \label{e:omega(A)=0}
  ||e^{tA}||\leq 1 \ , \ \forall t\geq 0 \quad \hbox{if and only if} \quad \omega(A)=0 \ . 
  \eea

\item[ii)] {\em Transient peak estimations: pseudospectral abscissa $\alpha_\epsilon(A)$
  and Kreiss constant ${\cal K}(A)$}

  The pseudospectrum provides a tool for a qualitative understanding of
  the transient process at intermediate times. In particular, a lower bound for the
  peak in the transient growth is given in terms of the pseudospectral abscissa
  $\alpha_\epsilon(A)$ for any $\epsilon>0$, namely
  \bea
  \label{e:etA_alpha}
  \sup_{t\geq 0}||e^{tA}|| \geq  \frac{\alpha_\epsilon(A)}{\epsilon} \quad, \quad
  \forall \epsilon>0 \ .
  \eea
  This means that large values of the resolvent in the right half-complex plane,
  with dependences in $\epsilon$ of the
  $\epsilon$-pseudospectrum $\sigma_\epsilon(A)$
  stronger  than linear (normal operator), lead to transient growths
  even if the  spectrum $\sigma(A)$ is well inside the stable left half-plane. 
  In practice~\cite{Schmi07,trefethen2005spectra},
  if the resolvent norm is found to be  $||R_A(z)||=||(z-A)^{-1}||=: S$
  for a $z\in\mathbb{C}$ with $\mathrm{Re}(z)= 1/\Delta T$ then, taking $\epsilon:=1/S$ and
  using the first characterization of the $\epsilon$-pseudospectrum in
  (\ref{e:pseudospectrum_def}), it holds  $\mathrm{Re}(z)\leq \alpha_\epsilon(A)$ and,
  from (\ref{e:etA_alpha}) 
  \bea
  \label{e:G(t)_MT}
  \sup_{t\geq 0}||e^{tA}|| \geq  \frac{\alpha_\epsilon(A)}{\epsilon} \geq
   \frac{\mathrm{Re}(z)}{\epsilon} = \frac{1}{\epsilon \; \Delta T} = \frac{S}{\Delta T} \ .
   \eea
   The quantity $\Delta T$ admits a dynamical interpretation, akin to that in modal
   analysis~\cite{trefethen2005spectra}:
   transients with growth bounded below by (\ref{e:G(t)_MT}) will actually happen in a time
   scale $\Delta T$.

   A tighter lower bound for the transient growth can be
   obtained by maximizing over $\epsilon>0$, leading to a lower bound in
  terms of the so-called  Kreiss constant ${\cal K}(A)$, namely
  \bea
  \label{e:Kreiss_constant_bound}
  \sup_{t\geq 0}||e^{tA}||  \geq {\cal K}(A) \ ,
  \eea
  where ${\cal K}(A)$ is defined --- and usefully characterised in terms of the resolvant norm--- as
  \bea
  \label{e:Kreiss_constant}
      {\cal K}(A) := \sup_{\epsilon > 0}\left\{ \frac{\alpha_\epsilon(A)}{\epsilon}\right\}
      = \sup_{\mathrm{Re}(z)>0}\{|\mathrm{Re}(z)|\cdot ||(z - A)^{-1}||\} \ .
      \eea
      Finer dynamical bounds in bounded time intervals $0<t\leq \tau$ can be found
      in~\cite{trefethen2005spectra,Davie05}.
      Finally, notice that in the case $\omega(A)=0$, we have
      that inequality (\ref{e:Kreiss_constant_bound}) is sharp, with
      $\sup_{t\geq 0}||e^{tA}||={\cal K}(A)=1$. This unity bound is attained at $t=0$
      for $\sup_{t\geq 0}||e^{tA}||$, as follows from inequality (\ref{e:omega(A)=0}),
      and in the limit $\epsilon\to\infty$ for ${\cal K}(A)=1$, as seen in
      (\ref{e:epsilon_to_infty}).

    \item[iii)] {\em Late time behaviour: spectral abscissa or spectral gap $\alpha(A)$}.
      
      The spectrum $\sigma(A)$ controls the late transient behaviour in both
      normal and non-normal dynamics. Specifically, the late time asymptotics are controlled
      by the spectral abscissa
  \bea
  \label{e:spectral_abscisa_late_behaviour}
  \lim_{t\to \infty} \frac{1}{t}\ln ||e^{tA}|| = \alpha(A) = \lim_{\epsilon \to 0} \alpha_\epsilon(A) \ ,
  \eea
  The characterization in terms of the pseudospectral abscissa $\alpha_\epsilon(A)$
  completes the interpolation role of the
  $\epsilon$-pseudospectrum from early ($\epsilon\to\infty$) to late time dynamics ($\epsilon\to 0$).
     For completitude, we note that a lower-bound counterpart of (\ref{e:eA_upper_bound})
   is given by 
  \bea
   \label{e:eA_lower_bound}
  ||e^{tA}||\geq e^{t\alpha(A)} \ , \ \forall t\geq 0 \ .
  \eea

\end{itemize}
  
\section{Pseudospectrum in the `close limit' approximation to BBH mergers}
\label{s:pseudospectrum_close-limit}
  Once we have discussed the role of $\epsilon$-pseudospectra in generic transients
  in non-normal linear dynamics, we explore the application of this
  framework to the BBH merger transient under the hypothesis of effective (non-normal)
  linear dynamics. Distinct physical mechanisms are in action along the different phases
  of the BBH dynamics---namely the initial inspiral, the merger phase itself and the
  late ringdown (cf. \cite{Schutz:2004uj}). In this setting, the  notion of $\epsilon$-pseudospectrum
  may offer a  unifying perspective of the BBH dynamics and waveform through its different
  regimes, interpolating from the late inspiral corresponding to $\epsilon\to \infty$ to the late ringdown
  at $\epsilon\to 0$, passing through from the transitional merger at intermediate values
  of $\epsilon$. We note the inverse relation between the time scale $t_{\mathrm{dyn}}$
  at each dynamical phase
  and $\epsilon$, namely
  \bea
  \label{e:t_1/epsilon}
  t_{\mathrm{dyn}}\sim \frac{1}{\epsilon} \ ,
  \eea
  that offers a qualitative tool to associate with particular structures/patterns in the 
  pseudospectrum ``topographic map'' of the resolvent norm (cf. e.g.  \cite{Jaramillo:2020tuu} for
  such a qualitative account of the pseudospectrum), with distinct dynamical phases
  in the BBH transient process.

  Specifically, some features of the BBH dynamics where one may explore the capability 
  of the $\epsilon$-pseudospectrum to offer qualitative/quantitative insights, include the following points:
  \begin{itemize}
  \item[i)] 
  Characterization of the transition from the late inspiral to the merger.
  \item[ii)] Estimation of the dominating peak at the merger.
  \item[iii)] Identification of time scales for the transition from merger
    phase to the ringdown phase.
  \item[iv)] Assessment of the possibility to account early/late dynamical behaviours
    in the merger transient process in terms of the specifics of the resulting final merged BH.
  \end{itemize}
  Exploring these points requires adopting a particular setting
  in which the pseudospectrum can be concretely realised.
  Very different scenarios, with distinct degrees of complexity, can be envisaged
  (cf. e.g. \cite{Jaramillo:2022mkh,JarKri22bv2}).
  In the following, as a first exploratory stage, we adopt an approach
  in the spirit of the
  ``close-limit'' approximation~\cite{Price:1994pm,Gleiser:1996yc,Khanna:1999mh} to BBH mergers.

  \subsection{BBH merger transient in the ``close-limit'' approach}
  As referred in the introduction, 
  the ``close-limit''
  approximation~\cite{Price:1994pm,Gleiser:1996yc,Khanna:1999mh} provides
  a model for the BBH coalescence dynamics, starting from the
  merger phase, in terms of the linear perturbations of the resulting merged and
  eventually stationary BH. In the long path leading to the successful numerical BBH
  mergers~\cite{Pre05,Pretorius:2007nq}, this model contributed by offering key insights into
  the emitted waveforms and  estimations of released energy. Interestingly, once the
  BBH problem is under numerical control and actual observational data are available,
  revisiting such models can be key to understand
  the physical mechanisms underlying the BBH problem.

  Our treatment here dwells in a ``close limit'' spirit, realised in the
  hyperboloidal setting to BH perturbations pioneered  by A. Zeng\i noglu \cite{Zenginoglu:2007jw,Zenginoglu:2011jz}
  and M. Ansorg and R.P. Macedo \cite{Macedo:2014bfa,Ansorg:2016ztf}, specifically in the form
  championed by  R.P. Macedo \cite{PanossoMacedo:2018hab,PanossoMacedo:2018gvw,PanossoMacedo:2020biw,Jaramillo:2020tuu}. For simplicity, in this first exploratory attempt
  we stay in a spherically symmetric Schwarzschild setting.
Following the discussion and the notation in  reference \cite{Jaramillo:2020tuu}, the relevant evolution equation is 
  \bea
  \label{e:wave_eq_1storder_intro}
  \left\{
  \begin{array}{l}
    \partial_\tau u = i L u \\
    u(\tau=0,x) = u_0(x)
\end{array}
  \right. \quad , \quad \hbox{with} \quad
   u(\tau, x) =
\begin{pmatrix}
  \phi(\tau, x) \\ \psi(\tau, x)
\end{pmatrix}  
\eea
where the $\tau$ time parameter is adapted to hyperboloidal slices and Eq. (\ref{e:wave_eq_1storder_intro}) is
the first-order reduction in time of the linear wave equation for $\phi$, with $\psi=\partial_\tau\phi$, where 
$\phi$ is an $(\ell, m)$ spherical harmonic mode of the axial/polar scalar master functions.
The operator $L$ writes as
\bea
\label{e:L_operator_intro}
L =\frac{1}{i}\!  \left(
  \begin{array}{c|c}
    0 & 1 \\ \hline L_1 & L_2
  \end{array}
  \right) \ ,
  \eea
where the explicit form of  operators $L_1$ and $L_2$ is given by
  \bea
  \label{e:L_1-L_2_intro}
 L_1 = \frac{1}{w(x)}\big(\partial_x\left(p(x)\partial_x\right) -
 \tilde{V}(x)\big) \; , \; 
 L_2 =
\frac{1}{w(x)}\big(2\gamma(x)\partial_x + \partial_x\gamma(x)\big) \ ,
\eea
where $x\in[a,b]$ is a compactified radial coordinate along the hyperboloidal slice
$\Sigma_\tau$, with $a$ and $b$
corresponding to the BH horizon and future null infinity $\scri^+$, the functions $p(x)$, $w(x)$ and
$\gamma(x)$ are fixed by the choice of slicing and compactification along the hyperboloid,
and $\tilde{V}(x) = V(x)/p(x)$, with $V(x)$ the effective potential corresponding to 
each gravitational parity (see \cite{Jaramillo:2020tuu} for details).  Choosing an `energy scalar product'
(cf. \cite{Jaramillo:2020tuu,Gasperin:2021kfv}) we can write
\bea
\label{eq_eff_en_inner_product}
   \langle u_1, u_2\rangle_{_E}  =
  \frac{1}{2} \hspace{-1mm}\int_{a}^{b}\hspace{-1mm}
  (w(x)\bar{\psi}_1 \psi_2 +
  p(x)\partial_x\bar{\phi}_1\partial_x\phi_2 +
  \tilde{V}(x)\bar{\phi}_1\phi_2) dx \ ,
\eea
that leads to the following expression for the formal adjoint $L^\dagger$ of $L$
\bea
\label{e:formal_adjoint}L^\dagger
=\frac{1}{i}\!  \left(
  \begin{array}{c|c}
    0 & 1 \\ \hline L_1 & L_2+ L_2^\partial
  \end{array}
  \right) \! \  ,
\eea
with  $L_2^\partial$ formally expressed in terms of Dirac-delta's supported at the boundaries
\bea
\label{e:L2_adjoint_pert} L_2^\partial = 2\frac{\gamma}{w} \bigg(
\delta(x-a)-\delta(x-b)\bigg) \  .
\eea
In the following, we apply the elements discussed in section \ref{s:transient_pseudospectrum}
to the transient case of BBH dynamics modelled in a (first-order)
`close limit' approximation  captured by 
Eqs. (\ref{e:wave_eq_1storder_intro})-(\ref{e:A_Adagger_BBH_closelimit}).

\subsection{BBH inspiral-merger transition}
\label{s:BBH_initial}
We start by exploring the possibility of describing the transition from the 
late inspiral phase to the merger phase by looking at the limit $\epsilon\to \infty$
of the pseudospectrum.
A necessary condition for making sense of such a description is having an initial positive
slope of the growth factor $||e^{i\tau L}||$, i.e. the positivity of its time derivative
at $\tau=0$, that according to the characterisation in Eq. (\ref{e:numerical_abscissa_early_behaviour})
translates into the
positivity of the numerical abscissa. The latter is indeed
characterised by the pseudospectrum in the $\epsilon\to \infty$ limit, as captured
in Eq. (\ref{e:epsilon_to_infty}). 
 
First, in order to match the notation in section \ref{s:Transients_and_pseudospectrum}
--- compare in particular the
evolution equations  (\ref{e:dyn_eq}) and (\ref{e:wave_eq_1storder_intro}) --- 
we identify  $A=iL$ that, at the spectral level, amounts to a $\pi/2$ rotation in the 
complex plane. In particular we can write (note $A^\dagger = -iL^\dagger$)
 \bea
\label{e:A_Adagger_BBH_closelimit}
  A = \left(
  \begin{array}{c|c}
    0 & 1 \\ \hline L_1 & L_2
    \end{array}
  \right)
\quad , \quad 
   A^\dagger =
    -\left(
  \begin{array}{c|c}
    0 & 1 \\ \hline L_1 & L_2+ L_2^\partial
  \end{array}
  \right)\ .
  \eea
  Using these expressions for $A$ and $A^\dagger$ we can calculate the
  symmetric part of the $A$ operator  
  \bea
\label{e:A+Adagger}
 \!\!\! \!\!\! \!\!\! \!\!\! \!\!\! \!\!\!\frac{1}{2}(A + A^\dagger) = \frac{1}{2}\left(
  \begin{array}{c|c}
    0 & 0 \\ \hline 0 &- L_2^\partial
  \end{array}
  \right)
  =
  \left(
  \begin{array}{c|c} 
    0 & 0 \\ \hline 0 &-\displaystyle \frac{\gamma}{w} \bigg(
\delta(x-a)-\delta(x-b)\bigg)
  \end{array}
  \right) \ ,
  \eea
where we have used the expression of $L_2^\partial$ in  Eq. (\ref{e:L2_adjoint_pert}). 
This remarkably simple expression corresponds to a diagonal operator (in particular,
we can consider the representation of the Dirac-delta in terms of limits of appropriate functions).
Therefore eigenvalues are given by the diagonal entries. In addition, and as a key consequence of
the outgoing nature of the asymptotic boundary conditions   
 ---realised in the hyperboloidal slicing that reaches future null infinity
 $\scri^+$ and in the (outgoing) transverse character of the slicing at the
  BH horizon--- it holds
  \bea
  \label{e:gamma<0}
  -\frac{\gamma(a)}{w(a)} < 0 \quad , \quad \frac{\gamma(b)}{w(b)} < 0 \ .
  \eea  
  From expressions (\ref{e:A+Adagger}) and (\ref{e:gamma<0}), and using  the spectral
  characterisation  (\ref{e:numerical_abscissa_characterization}) of the 
numerical abscissa, in terms of the supremum
of the real part of the spectrum of $\frac{1}{2}\left(A + A^\dagger\right)$,
it follows
  \bea
\label{e:omega(L)=0}
  \omega(L) = 0 \ .
  \eea
We notice that this result is essentially independent of the chosen foliation,
only depending on asymptotic boundary values, actually
the part responsible of the loss of selfadjointness of the operator
as a consequence of the field leaking through the boundaries~\cite{Jaramillo:2020tuu,Gasperin:2021kfv}.
Crucially, we notice that this result does not depend on the specific 
nature of  the potential $V(x)$.

The blunt  consequence of Eq. (\ref{e:omega(L)=0}) is that no
peak can happen in the later dynamical evolution. Therefore this simple model
---namely, a first-order perturbation `close-limit' BBH model---
cannot account for the transition from the late inspiral to the merger phase.
On the other hand, the fact that $\omega(L)$ is not negative, but 
strictly vanishing, entails a non-trivial consequence: 
 $\omega(L) = 0$ is consistent with a description in terms
of non-normal linear dynamics starting 
at the merger peak. This is akin with the results in
\cite{Giesler:2019uxc,Isi:2019aib,Capano:2021etf} and opens  
the possibility of exploring the full BBH dynamics from a
non-modal analysis/pseudospectrum perspective
right from the merger peak, as in the original close-limit approximation spirit.

  \subsection{Kreiss-constant characterisation of the BBH merger waveform maximum}
  \label{e:Kreiss-BH}
  Applying the discussion of the Kreiss constant ${\cal K}(L)$ in section
  \ref{s:transient_pseudospectrum} to BBH mergers, one would like
  to assess the possibility of estimating the merger 
peak from ${\cal K}(L)$. Unfortunately, this expectation does not stand in the
close-limit approximation to BBH mergers. From the vanishing of the numerical 
abscissa in Eq. (\ref{e:omega(L)=0}), we can conclude from  (\ref{e:omega(A)=0})
that $e^{itL}$ is contractive and, as discussed after Eq. (\ref{e:Kreiss_constant}),
it holds ${\cal K}(L)=1$.
We conclude that finer models than the first-order close-limit approximation are needed 
to explore the use of the Kreiss constant to estimate
the merger peak in a non-normal linear description of BBH mergers.

With these more general  BBH settings in mind, let us make some general considerations
on the Kreiss constant and, more generically, on the $\epsilon$-pseudospectrum 
with intermediate $\epsilon$'s:
\begin{itemize}
\item[i)] {\em Kreiss constant and pseudospectrum}. 
  For the sake of clarity, we translate the expression of ${\cal K}(A)$ in terms of $L$, namely
  performing a clockwise $\pi/2$-rotation 
in Eq. (\ref{e:Kreiss_constant}) we get
  \bea
  {\cal K}(L) = \sup_{\mathrm{Im}(z)<0}\{|\mathrm{Im}(z)|\cdot ||(L-zI)^{-1}||\} \ .
  \eea
We notice that the evaluation of the Kreiss constant is straightforward 
(and comes ``for free'') during the process of calculating the pseudospectrum numerically, 
since the latter consists precisely in evaluating the  resolvent's norm $||(L-zI)^{-1}||$ at each
point $z\in\mathbb{C}$.

\item[ii)] {\em Kreiss constant and gravitational wave strain at the merger peak}. The Kreiss constant
  provides a lower bound estimate of the growth factor $G(t)$ in Eq. (\ref{e:G(t)}),
  namely the ratio between the amplitude of the propagating field $\phi(t)$ and an initial reference value. 
  This dimensionless quantity should provide an estimation of the gravitational wave strain $h$ 
  \bea
  h \gtrsim {\cal K} \ ,
  \eea

\item[iii)] {\em Pseudospectral abscissa timescale}. The patterns appearing in the pseudospectrum
  inform about phenomena happening at different time scales. The discussion around
  expressions (\ref{e:G(t)_MT})
  can be recast in the BBH setting by introducing a $\epsilon$-pseudospectrum timescale as
  \bea
  \label{e:tepsilon}
   t_{\mathrm{\epsilon}} :=\frac{1}{\alpha_\epsilon(L)}
\eea
where now $\alpha_\epsilon(L) = \sup|\mathrm{Im}(\sigma^\epsilon(L)|$. In other words,
dynamical phenomena associated with a given $\epsilon$-pseudospectrum pattern manifests
at a time scale $t_\mathrm{\epsilon}$. The inequality
\bea
\label{e:tepsilon_epsilon}
t_{\mathrm{\epsilon}} \cdot \epsilon \geq \left( {\cal K}(L)\right)^{-1} \ ,
\eea
following from (\ref{e:Kreiss_constant}) gives a more precise content~\footnote{
  From inequality (\ref{e:tepsilon_epsilon}) we can write $t_{\mathrm{\epsilon}}$ as finer
  quantity than $t_{\mathrm{dyn}}$ in inequality (\ref{e:t_1/epsilon}), for estimating dynamical times
    \bea
    \label{e:T_epsilon_K}
    t_{\mathrm{\epsilon}} \gtrsim ({\cal K}(L))^{-1} \cdot \frac{1}{\epsilon} = ({\cal K}(L))^{-1} \cdot t_\mathrm{dyn} \ .
    \eea
    Later, we will revisit the interpretation of
    $({\cal K}(L))^{-1}$ as a ``minimal action'' $h_{\cal K}$ in the ``uncertainty relation'' (\ref{e:uncertainty}).} to the  timescale estimate in
(\ref{e:t_1/epsilon}).

\end{itemize}

\subsection{Spectral abscissa: the question of QNM spectral (in)stability}
\label{e:BBH_ringdown}
The late time decay is controlled by the spectrum $\sigma(L)$ of $L$, namely the
$\epsilon$-pseudospectrum $\sigma^\epsilon(L)$ in the limit $\epsilon\to 0$. This
holds both for normal and non-normal dynamics.  Specifically, at sufficiently late times 
the spectral abscissa $\alpha(L)$ controls the decay. In the asymptotically
flat  BH case the very late decay is driven by tails corresponding to the 
continuous part of the spectrum. Before the tails take over, the late decay
is controlled by the imaginary part of the fundamental QNM. As discussed
above, for earlier times the spectrum does not capture the dynamics in the
non-normal case and intermediate $\epsilon$-pseudospectra are needed.
A natural question concerns the role of QNM overtones in the BBH transient 
and ringdown dynamics. 

\setcounter{footnote}{0}

The latter question, crucial in BH spectroscopy~\cite{Berti:2005ys,Dreyer:2003bv,Baibhav_2018,Ota:2019bzl,Isi:2019aib,Giesler:2019uxc,Isi:2020tac,Cabero:2019zyt,Maggio:2020jml,Ota:2021ypb},
becomes particularly relevant in the context of the recent
discussion of a possible instability of the QNM spectrum~\cite{Jaramillo:2020tuu,Jaramillo:2021tmt,Gasperin:2021kfv}.
The BH pseudospectrum (in the `energy norm') shows non-trivial patterns 
extending far from the spectrum. This is a signature of spectral instability.
The analysis in refs. \cite{Jaramillo:2020tuu,Jaramillo:2021tmt,Gasperin:2021kfv}
suggests the presence of a  QNM overtone instability under high-wave number 
(small scale) perturbations that, crucially,  leaves the fundamental QNM (and therefore the
 QNM spectral abscissa' $\alpha (L)$) unchanged. Another kind of ('flea on the elephant') instability affecting the 
 fundamental QNM has been proposed in \cite{Cheung:2021bol}~\footnote{Both kind of instabilities are ultimately encoded
   in the pseudospectrum,
but respond to different mechanisms (namely, from a semiclassical
perspective, they are associated with distinct `closed' phase space trajectories
\cite{zworski2017mathematical,BinZwo}).}.
However, these QNM instability proposals must address the fact that the global 
qualitative aspects of the time evolution after the BBH merger
seems quite insensitive to such spectral instabilities.

\subsection{Spectral and time-domain perspectives: QNMs and dynamics.}
We  discuss now some points aiming at reconciling these seemingly 
conflicting spectral and dynamical perspectives and, more generally,
at shedding light on QNMs and dynamics: 
\begin{itemize}

\item[i)] {\em Pseudospectrum universality and early-intermediate time evolution}. 
  Before the late ringdown, the average qualitative aspects of the transient dynamics are
  not imprinted by possible perturbations of the QNM spectrum, in particular by perturbed QNM overtones.
  This fact is not surprising since, as discussed above, dynamics at early and intermediate
  timescales are not controlled by the spectrum ($\epsilon=0$) but by $\epsilon$-pseudospectra
  with large/intermediate $\epsilon$ values. 
  In this setting, the following fact is crucial:
  \begin{center}
  {\em The pseudospectrum (asymptotic) structure
    is universal, shared by unperturbed and perturbed BH potentials, therefore
    leading to qualitatively similar early dynamics.}
  \end{center}
  This BH pseudospectrum feature has been described in \cite{Jaramillo:2020tuu} and further explored in
  \cite{Destounis:2021lum}. As a consequence, the qualitative dynamics at early and intermediate
  timescales after the BBH merger are universal, therefore qualitatively independent of 
  non-perturbed BH QNMs or particular instances of perturbed (Nollert-Price-like) open BH QNM branches.

\item[ii) ]{\em Pseudospectra regularization and evolution timescales}. We can refine the previous 
  notion of universality of pseudospectra, specifically in the context of small scale perturbations.
  A remarkable effect of (random) small-scale perturbations of `energy size' $\epsilon_o$  
  is the `flattening' of the $\epsilon$-pseudospectra patterns with 
$\epsilon\leq \epsilon_o$. This is illustrated in Fig. \ref{fig:Random_regularization}, showing the
  pseudospectrum corresponding to the P\"oschl-Teller potential and to different perturbations of this
  potential with increasing $||\delta \tilde{V}||_E := \delta L = \epsilon_o$ (figures are taken from
  \cite{Jaramillo:2020tuu},
  where a discussion is presented in further detail). 
  This phenomenon is a consequence of the regularization effect under random (small scale) perturbations on
  the analytic structure of the resolvent  $R_L(z)=(L - z I)^{-1}$~\cite{hager05,Hager06a,Hager06b,HagSjo06,Borde08,BorSjo10,Borde11,Borde13,Vogel16,NonVog18,Sjostrand2019}. The crucial fact in our context is that the (contour lines of)
  $\epsilon$-pseudospectra  
with $\epsilon>\epsilon_o$ remain strictly unaltered (cf. Fig.  \ref{fig:Random_regularization}).
In other words, dynamics at earlier times than $t_{\mathrm{\epsilon_o}}$, corresponding to $\epsilon>\epsilon_0$
(cf. expressions (\ref{e:tepsilon}) and
(\ref{e:tepsilon_epsilon})) are not affected by small-scale perturbations with ``energy size'' $\epsilon < \epsilon_o$,
dynamical deviations effects only appearing at times later than  $t_{\mathrm{\epsilon_o}}$ corresponding to
the `flattened' $\epsilon$-pseudospectra with $\epsilon<\epsilon_o$~\footnote{Note that this is
  consistent with the eventual dominating role of the spectrum ($\epsilon\to 0$) at very late times.}.
This discussion has a interesting reinterpretation in terms of a formal `time-energy uncertainty relation':
denoting the timescale $t_{\mathrm{\epsilon}}$ and the perturbation `energy size' $\epsilon$, respectively
by $\Delta t$ and $\Delta E$, and $\left( {\cal K}(L)\right)^{-1}$ as $h_{\cal K}$, we can write inequality
(\ref{e:tepsilon_epsilon}) as~\footnote{If we want to push further this formal heuristic interpretation,
  the peak at the merger would correspond to a ``coherent state''
  saturating the uncertainty inequality $\Delta t \cdot \Delta E\gtrsim  h_{\cal K} = \left( {\cal K}(L)\right)^{-1}$.}
\bea
\label{e:uncertainty}
\Delta t \cdot \Delta E \geq h_{\cal K} \ .
\eea
This should be read as follows: small-scale perturbations with energy of order $\Delta E$ only
affect transient dynamics in timecales larger than $\Delta t$'s constrained by inequality (\ref{e:uncertainty}).

\begin{figure}[t!]
\begin{center}
\includegraphics[width=6.2cm]{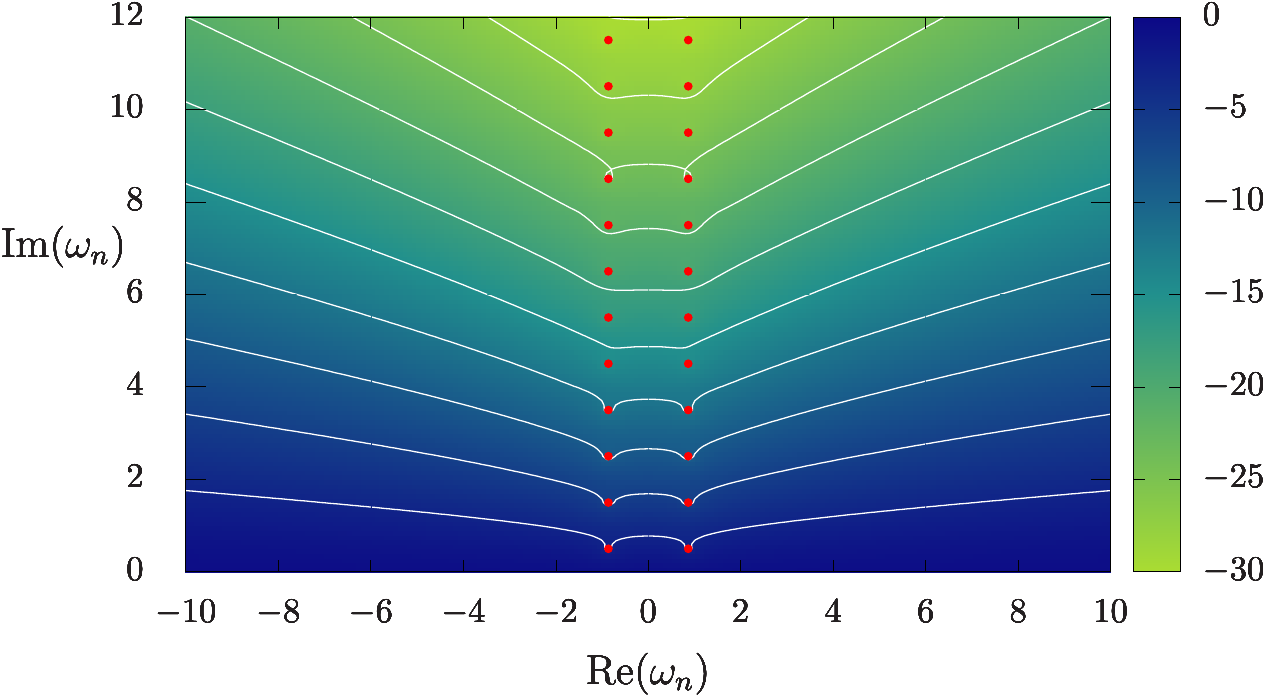}
\includegraphics[width=6.2cm]{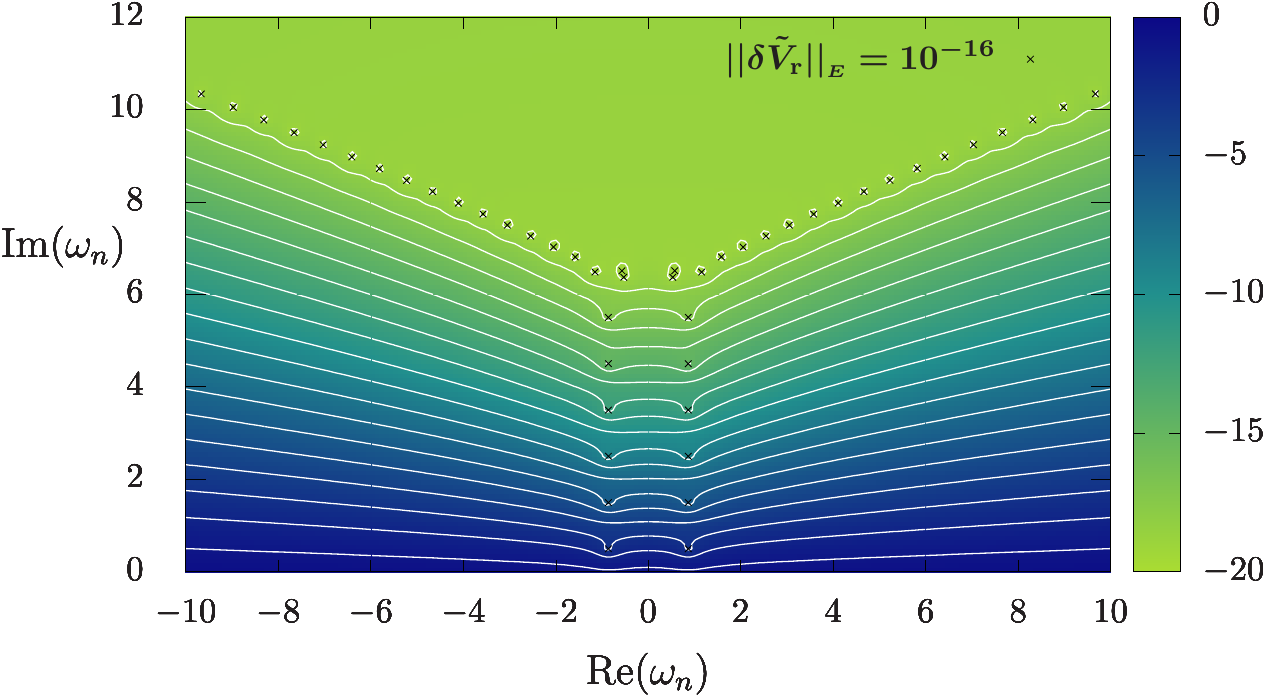}
\includegraphics[width=6.2cm]{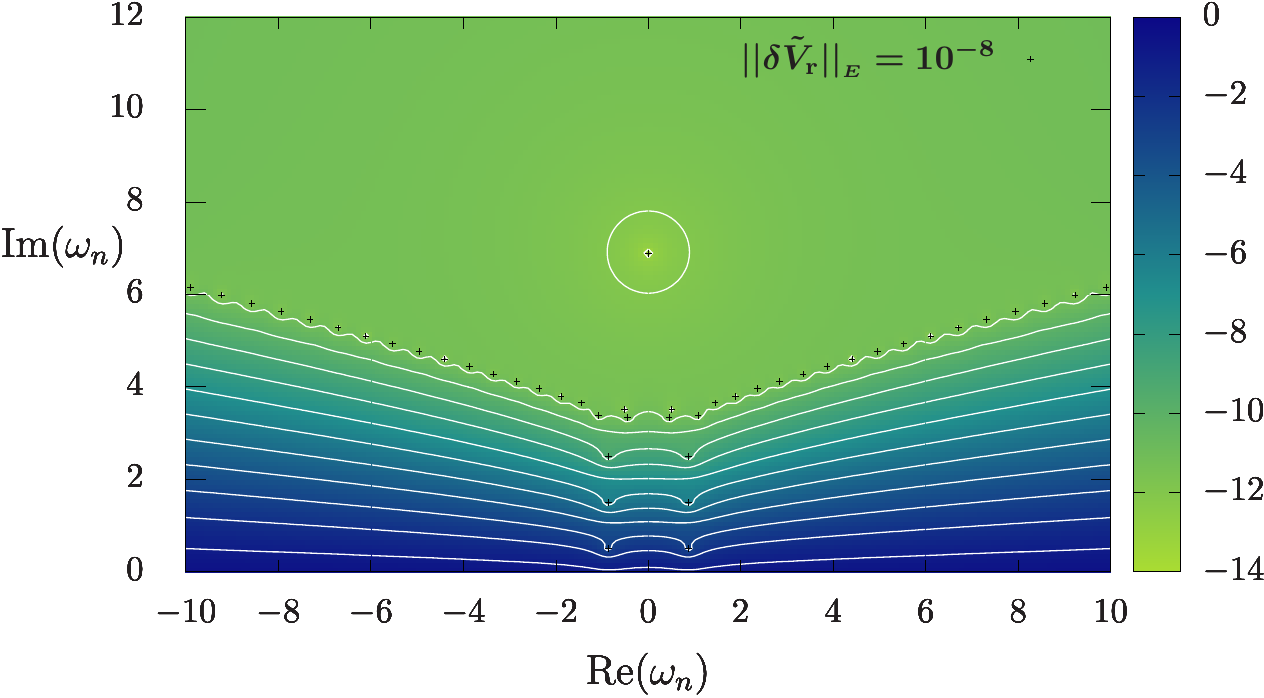}
\includegraphics[width=6.2cm]{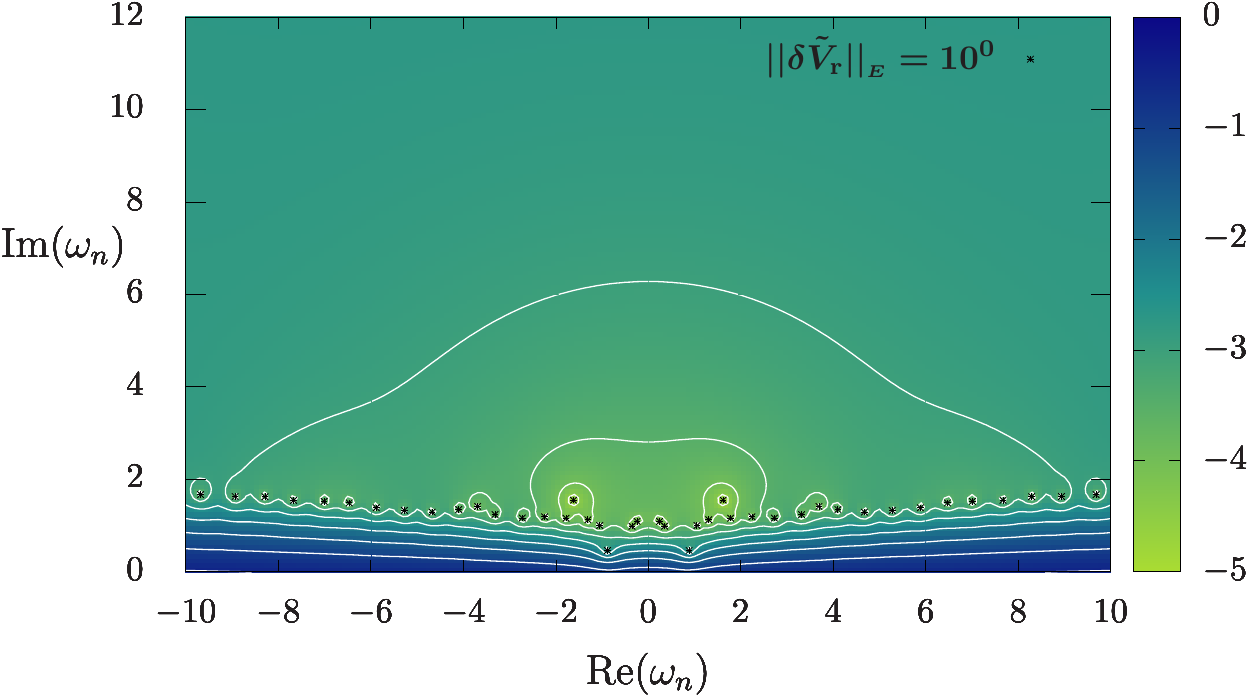}
\end{center}
\caption{Pseudospectrum regularization and evolution timescales. White lines correspond to
  contour lines of $\epsilon$-pseudospectra, with $\epsilon$ growing towards the dark green
  regions (i.e. downwards in the figure). High wave-number (random) perturbations of size
  $\epsilon_o = ||\delta\tilde{V}||_E$
  flatten the pseudospectrum (namely, regularize the resolvent $R_L(z)$) for $\epsilon<\epsilon_o$,
  therefore modifying the dynamics
  for $t_\epsilon>t_{\epsilon_o}$. Pseudospectra lines for $\epsilon>\epsilon_o$ remain however unaltered,
indicating that early dynamics with  $t_\epsilon<t_{\epsilon_o}$ are not affected by these perturbations.}
\label{fig:Random_regularization}
\end{figure}

\item[iii)] {\em Pseudospectrum regularization and $\epsilon$-dual QNM expansions.} In ref.
 \cite{Gasperin:2021kfv} the notion of `$\epsilon$-dual QNM expansions' is introduced, based 
 on the `stability' (in the energy norm) of the time evolution, in contrast with the spectral QNM instability.
In brief, considering the evolution operator $L(x)$ in a stationary spacetime
and the {\em stationary} perturbation $\delta L(x)$ giving rise to the perturbed
evolution operator $L^\epsilon = L + \epsilon \delta L$ (with $||\delta L||=1$), 
the respective  (right-)spectral problems obtained under Fourier transform of (\ref{e:wave_eq_1storder_intro})
in time are
\bea
\label{e:QNM_exact_pert}
L \hat{v}_n = \omega_n  \hat{v}_n \quad , \quad L^\epsilon \hat{v}^\epsilon_n(x) =\omega^\epsilon_n \hat{v}^\epsilon_n(x)  \ .
\eea
Then, it is shown \cite{Gasperin:2021kfv} that the corresponding unperturbed and perturbed
evolution fields, $u(\tau,x) $ and $u^\epsilon(\tau,x)$, admit  natural
(Keldysh) QNM-resonant asymptotic expansions in the respective unperturbed and perturbed QNM frequencies and
eigenfunctions
\begin{equation}
\label{e:resonant_expansions}
u(\tau,x) \sim  \sum_n e^{i\omega_n \tau} a_n \hat{v}_n(x) \quad, \quad
u^\epsilon(\tau,x) \sim 
\sum_n e^{i\omega^\epsilon_n \tau} a^\epsilon_n \hat{v}^\epsilon_n(x) \ .
\end{equation}
From the (energy norm) stability  of the time evolution,
it holds 
$||u - u^\epsilon||_{_E} \ \lesssim C \epsilon$, so we can write 
$u^\epsilon(\tau,x)\sim u(\tau,x) +O(\epsilon)$. The associated QNM expansions cannot
be distinguished at order $\epsilon$ and are therefore equivalent at order $\epsilon$,
something  we write~\cite{Gasperin:2021kfv}
\bea
   \label{e:epsilon-dual_expansions}
   \sum_n e^{i\omega^\epsilon_n\tau} a^\epsilon_n
   \hat{v}^\epsilon_n(x) \stackrel{\epsilon}{\sim} \sum_n
   e^{i\omega_n\tau} a_n \hat{v}_n(x) \ .  
\eea 
Both expansions are indistinguishable up to errors of order $\epsilon$, providing two
alternative descriptions of the evolved field at this order~\footnote{We note
    that QNM-expansions (\ref{e:epsilon-dual_expansions}) can be in principle completely 
    disentangled if full control of the initial data is available, fixing the respective
    coefficients $a_n$ and $a^\epsilon_n$, as illustrated 
    in \cite{Jaramillo:2021tmt}. An in-depth study of this question is required to
    assess the challenges for BH spectroscopy due to plausible degeneracy issues in the data analysis problem.}.
This picture is consistent with the
above-discussed regularization of  the pseudospectra by order $\epsilon_o$ perturbations:
for early times (with $\epsilon \geq \epsilon_o$)
the pseudospectra and associated time evolutions are indistinguishable, whereas
for late times (with $\epsilon \leq \epsilon_o$) the description
may differ. In other words, $\epsilon$-dual QNM expansions correspond
to equivalent resonant expansions for early times, associated with the unperturbed part of the
pseudospectrum under order-$\epsilon_o$ perturbations.

\item[iv)] {\em Non-stationary perturbations and time averaging}. The non-perturbed
and perturbed QNM problems
in Eq. (\ref{e:QNM_exact_pert}) assume the stationarity of the operator $L$ and the
perturbation $\delta L$. The same holds for the problems studied in 
\cite{Jaramillo:2020tuu,Jaramillo:2021tmt,Gasperin:2021kfv,Destounis:2021lum} and, more generally, the stationarity
hypothesis pervades the standard  approach to the very definition of QNMs~\footnote{However, for a
  time-dependent resonance theory by Soffer and Weinstein, see~\cite{SofWei98} (see also
  Lax and  Phillips \cite{LaxPhi89}).}.
 Whereas the stationarity of $L$ is a reasonable hypothesis due to the
 uniqueness BH theorems, the stationarity of the perturbation $\delta L$ is a strong physical hypothesis.
  If we rather consider the evolution problem under a time-dependent perturbation
  $\delta L(\tau,x)$
  \bea
  \label{e:perturbed problem_t}
  \partial_\tau u^\epsilon(\tau,x) = \big(L(x) + \epsilon \delta L(\tau,x)\big)u^\epsilon(\tau,x) \ ,
  \eea
  a natural approximation approach 
 (in particular, for oscillating $\delta L(\tau,x)$'s of
  period $T$) is to adopt an averaging method (e.g. \cite{Perko69,murdock1999perturbations})
  to set a time-independent problem
  \bea
  \!\!\!\!\!\!\!\!\!\!\!\!\!\!\!\!\!\! \partial_\tau \bar{u}^\epsilon(\tau,x) =  \big(L(x) + \epsilon \overline{\delta L}(x)\big)\bar{u}^\epsilon(\tau,x)  
  \ \ , \ \
  \overline{\delta L}(x) = \frac{1}{T}\int_0^T \delta L(t,x) d\tau \ ,
  \eea
  where the difference between $u^\epsilon$ and the averaged $\bar{u}^\epsilon$
  is bounded in the time $\tau$, by
  \bea
  \label{e:averaged_error}
  ||u^\epsilon - \bar{u}^\epsilon||(\tau) < C \epsilon \ ,
  \eea
  with $\tau$ in the interval $\tau\!\in]0, \frac{\tau_0}{\epsilon}[$, for some constants $C$ and $\tau_0$.
  This has direct implications in our spectral BH QNM spectral instability problem.
  In particular, if we consider an oscillating  perturbation averaging to zero
  in time, i.e.  
  $\overline{\delta L}(x) = 0$ (as it is indeed reasonable to assume for numerical noise
  in BBH simulations), the averaged signal $\bar{u}^\epsilon(\tau,x)$ corresponds
  actually to the non-perturbed signal  $u(\tau,x)$ in the non-perturbed
  problem. Only if we go to time-samplings below the period $T$ of the oscillation,
  i.e. if we consider $\tau_0/\epsilon$ sufficiently small and therefore with an $\epsilon$
  in (\ref{e:averaged_error}) large enough,  would the averaged signal depart from the
  unperturbed one. This mechanism seems to be adequate to explain the absence of observed
  departures from the expected, non-perturbed results, in BBH {\em numerical} simulations.
  Regarding actual astrophysical observational data,
  if the sampling of the time-series for the ringdown signal is coarser that the
  period of the (potential) astrophysical perturbations and the latter average to zero
  over such timescales, again, we would retrieve unperturbed QNMs from the ringdown signal.
  This poses a challenging technological problem to retrieve fastly oscillating
  astrophysical sources.

\end{itemize}

\section{A unifying effective linear-dynamics approach to BBH dynamical phases}
\label{e:conclusions}
In these notes we have proposed the pseudospectrum notion and the related non-modal
analysis as a framework to study the BBH merger waveform as a transient phenomenon,
under the (strong) hypothesis of effective BBH non-normal linear dynamics.
This approach provides a unified scheme to study the different
phases of the BBH waveform in terms of $\epsilon$-pseudospectra, the (late)
inspiral phase corresponding to the limit $\epsilon\to\infty$, the final ringdown
characterised by $\epsilon\to 0$ and the actual merger by intermediate
values of $\epsilon$. In particular, qualitative insights gained
from such unified pseudospectrum perspective may provide clues in
the effort to understand the simplicity and universality
of BBH merger waveforms~\cite{Jaramillo:2022mkh,JarKri22bv2}.

Specifically, given the infinitesimal dynamical time generator $L$,
we have introduced: i) the numerical abscissa $\omega(L)$,
controlling the triggering of the dynamical transient, 
ii) the pseudospectral abscissa $\alpha_\epsilon(L)$ and the Kreiss constant
${\cal K}(L)$ offering estimates of the transient peak,
and iii) the spectral abscissa  $\alpha(L)$ (or spectral gap)
controlling the late time behaviour. We have endowed the $\epsilon$ parameter
in $\epsilon$-pseudospectra with an interpretation as an inverse dynamical
timescale, namely $t_{\mathrm{dyn}}\sim t_\epsilon\sim 1/\epsilon$, in such a way that
non-trivial patterns in the $\epsilon$-contour lines of the pseudospectrum
qualitatively inform of dynamical processes occurring at the $t_\epsilon$ timescale.
In particular, as a consequence of the asymptotic universality of the
pseudospectrum at large $\epsilon$'s, we have concluded that early BBH
merger waveforms are largely independent of possible (environmental) perturbations
in the BBH dynamics, deviations only appearing at late times
controlled by the spectrum, i.e. encoded in possibly perturbed QNMs.

Concerning the latter context of perturbations in (after merger) BBH dynamics,
namely the question of BH QNM spectral instability,
we have discussed the following refinements:
i) perturbations with `energy' of order $\Delta E$ only
affect post-merger dynamics in timecales larger than a $\Delta t$
constrained by $\Delta E\cdot \Delta t \geq ({\cal K}(L))^{-1}$,
ii) `$\epsilon$-dual' QNM expansions associated with non-perturbed and
perturbed QNMs provide equivalent resonant expansions for early times $t\leq t_\epsilon$,
and iii) non-stationary perturbations averaging to zero in a timescale $T$
render {\em exactly} the non-perturbed BH QNMs (in spite of the BH QNM spectral instability
phenomenon) if the sampling-time of the signal is larger than $T$,   
QNM instabilities only becoming apparent for finer time samplings.
The latter point is crucial for assessing time-varying BH perturbations.

The present notes must be understood as an introductory invitation to the subject,
aiming at prompting a deeper and systematic application of non-modal
analysis~(e.g. \cite{TreTreRed93,Trefe97,trefethen2005spectra,EmbTre_webpage,Sjost03,Davie05,Davie07,Schmi07,Sjostrand2019}), well-established in the hydrodynamics context, to
the gravitational setting and, in particular, to the BBH problem.
As an illustration, the non-modal analysis tools specifically discussed here
have been implemented in the particular instance
provided by the `close limit' approximation to BBH dynamics. Specifically,
a fully explicit calculation of the numerical abscissa $\omega(L)$
(and also of the Kreiss constant ${\cal K}(L)$) has been presented,
making use of a compactified hyperboloidal approach to linear
dynamics over a Schwarzschild background.
The conclusion is that the BBH merger peak cannot be accounted in
such `close limit' approximation,
therefore this particular Ansatz for the  effective BBH linear dynamics
does not provide a good modelling of the dynamical
transient from the  inspiral to the merger phases.
At the first-order in perturbation theory, this follows from the
vanishing of the numerical abscissa, namely $\omega(L)=0$. At the
same time, this very result supports the validity of the `close limit'
in the dynamics starting at the peak of the merger.
Even if we go to second-order perturbation theory,
the `close limit' approximation does not provide a good account of the transient
peak, as shown in the~\ref{a:pseudoresonances}, where
the pseudospectrum is used to assess the possibility of understanding
the merger peak in terms of a so-called `pseudo-resonance', but without success
(the `pseudo-resonance' concept can be however useful on other (ultra-)compact object
settings, as explored in~\cite{BoyCarDes22}). The bottomline
of this `close-limit' approximation analysis is that a refined
model is needed to explore the effective linear dynamics hypothesis,
the key point being ---as indicated in section~\ref{s:eff_linear}---
the correct identification of the relevant background dynamics. This
is precisely the goal in~\cite{JarKri22bv2}.

Addressing this latter point from a more general perspective, in these notes 
we have discussed the possibility of studying
the BBH merger process as a dynamical transient described
by the (non-normal) linear dynamics for some (fast) degrees of freedom $u$
determined by the initial data
problem (\ref{e:dyn_eq}) or, in the notation
in \cite{Jaramillo:2020tuu}, by the equation (\ref{e:wave_eq_1storder_intro}).
As indicated in the introductory section~\ref{s:eff_linear}, this is a bold
assumption in the setting of a non-linear theory such as general relativity.
Even more, it is in manifest tension with sound approaches to the BBH merger
problem that vindicate precisely the key role of non-linearities
\cite{Sberna:2021eui}. Our approach in this sense is an `agnostic'
one open to intermediate effective treatments integrating both linear and
genuinely non-linear mechanisms, each one addressing distinct but
concurrent phenomena. An example of this is provided in the study of
hydrodynamical instability  and
 turbulence, where linear non-normal transient growths
 can indeed act as the seed triggering the full development of the non-linear
 turbulent regime~\cite{TreTreRed93,Schmi07}, in a kind of `bootstrapping'
 mechanism. Specifically in our gravitational setting,
 the present discussion has focused on the effective
 linear dynamics of fields $u$ propagating
 on a (fixed) background determined by the dynamical time operator $L$.
 A natural further step would consist in endowing $L$ with genuine (effective) non-linear 
 dynamics corresponding to (slow) background degrees of freedom, very much in
 the spirit of `wave-mean flow' approaches in hydrodynamics~\cite{buhler_2014}.
 In this context, the present pseudospectrum transient discussion
 is a part of a more ambitious program aiming at
 understanding the qualitative mechanisms of BBH dynamics, in particular the
 simplicity and universality of BBH merger waveforms~\cite{Jaramillo:2022mkh,JarKri22bv2}.

\section*{Acknowledgments}

The author would like to thank Badri Krishnan and Carlos F. Sopuerta for
discussions on the binary black hole problem. I would also like to thank Valentin Boyanov,
Vitor Cardoso, Kyriakos Destounis and Rodrigo P. Macedo for discussions on the pseudospectrum
and transients, Emanuele Berti for his insightful account on the pseudospectrum in BHs,
 Antonin Coutant for many discussions on `wave-mean field' theory in hydrodynamics and
Luis Lehner for sharing his understanding of non-linearities in the BBH ringdown.
More generally, I would also like to warmly thank  Edgar Gasper\'\i n, Rodrigo P. Macedo, Oscar Meneses-Rojas,
R\'emi M. Mokdad, Lamis Al Sheikh and Johannes Sj\"ostrand for continuous discussions on the pseudospectrum,
as well as Piotr Bizo\'n and Oscar Reula for their generous share of insights in PDE dynamics.
 This work was supported by the French ``Investissements d'Avenir'' program through
 project ISITE-BFC (ANR-15-IDEX-03), the ANR ``Quantum Fields interacting with Geometry'' (QFG) project
 (ANR-20-CE40-0018-02),  the EIPHI Graduate School (ANR-17-EURE-0002) and  
  the Spanish FIS2017-86497-C2-1 project (with FEDER contribution).
\bigskip 

\appendix

\section{Pseudospectrum and pseudo-resonances: assessing the BBH merger peak}
\label{a:pseudoresonances}

In the spirit of integrating both linear and genuinely non-linear mechanisms
in BBH dynamics,
we consider in this appendix another important aspect of the pseudospectrum
 that may be of relevance when incorporating general relativistic
 non-linear effects in a higher-order perturbation scheme~\footnote{We can identify
   at least three settings in which the pseudospectrum plays an important role
   in the analysis of non-normal operators and their associated dynamics
   (cf. \cite{TreTreRed93}): i) the assessment of spectral instability, ii) the
   study of transient growths in purely linear (non-normal) dynamics, and iii) the
   study of so-called pseudo-resonances. In the BBH context we have addressed the
   aspect i) in \cite{Jaramillo:2020tuu,Jaramillo:2021tmt,Gasperin:2021kfv,Destounis:2021lum},
   whereas point ii) is the subject of the main text in these notes, namely sections
   \ref{s:Transients_and_pseudospectrum} and \ref{s:pseudospectrum_close-limit}.
   This appendix addresses the aspect iii) corresponding  to pseudo-resonances.}. 
 Our non-modal analysis discussion has been based
 on the capability of the pseudospectrum to explain transients in
 non-normal initial value problems, without external forcing. However, the
 pseudospectrum is also relevant to account for  dynamical effects
 when the non-normal linear system is driven by an external force, leading to the
 notion of `pseudo-resonance': 
 enhanced resonant responses to external forces, not related
 to the QNM spectrum, occurring when  $\epsilon$-pseudospectra reach large values
 far from the spectrum.
 In our BBH context, the external forcing is precisely given by
 second-order perturbation theory.

 General relativity perturbation theory is a delicate subject plenty
 of subtle conceptual and technical points.
 Here we dwell at a purely formal level, interested in the resulting
 hierarchy of linear equations at different orders, with a focus on
 the second order.
 Following~\cite{Isaac68a,CunPriMon80} (see also \cite{Miller:2016hjv,Pound:2015fma}
 for the specific notation), we expand the metric $g_{ab}$ to second order as
 \bea
 g_{ab} = g^{(0)}_{ab} + \epsilon h^{(1)}_{ab} +  \epsilon^2 h^{(2)}_{ab} + O(\epsilon^3) \ .
 \eea
 Adopting gauges such that perturbed (vacuum) Einstein equations
 to second order write as
 \bea
 \label{e:GR_pert}
 \delta G_{ab} \cdot h^{(1)} &=& 0 \nn \\
 \delta G_{ab} \cdot h^{(2)} &=& \delta^2G_{ab}[h^{(1)}, h^{(1)}] \ ,
 \eea
 where $\delta G_{ab}$ is the linearized Einstein tensor
 evaluated on $g^{(0)}_{ab}$ acting linearly on $h^{(1)}_{ab}$
 and $h^{(2)}_{ab}$, respectively at first and second-perturbation order
 (we use the notation ``$\cdot$'' to emphasize the linear action).
 Then $\delta^2G_{ab}[h^{(1)}, h^{(1)}]$ is 
 the second-order variation of the Einstein tensor evaluated on the first-order
 metric perturbation (having a formal structure $\delta^2G\sim
 h^{(1)}\partial^2 h^{(1)} + \partial h^{(1)}\partial h^{(1)}$).
 In summary, we have a hierarchy of linear equations crucially sharing {\em the same}
 left-hand side, given by
 differential linear operator $\delta G_{ab}$ corresponding to the non-perturbed
 metric, and
 with right-hand side sources determined at each order by lower-order
 solutions~\footnote{I thank R.P. Macedo for bringing
   my attention to this important structural aspect of general relativity
   perturbation  equations, crucial in the application of a non-modal pseudospectrum analysis
   to the non-perturbed operator $\delta G_{ab}$.}. In vacuum, the
 first-order in this hierarchy is a homogeneous equation.

 Having sketched the general relativistic perturbation setting, we revisit now the
 close-limit approximation to BBH mergers.
 Specifically, in section \ref{s:pseudospectrum_close-limit} of the main text we have addressed the close-limit
 approximation through its formulation relying on the general relativity perturbation
 theory at  first-order~\cite{Price:1994pm}. We have seen that this approximation
 does not account for the transition from
 the late inspiral to the merger phase and, in particular, for the
 merger peak. However, the close-limit approximation can be upgraded
 to an improved version at second-order in perturbation theory~\cite{Gleiser:1996yc}.
 Specifically, Eqs. (\ref{e:GR_pert}) can be cast as~\cite{Gleiser:1996yc,CunPriMon80}
 \bea
 \label{e:close-limit_2ndorder}
  \left(\partial_\tau -  i L\right) u^{(1)} &=& 0 \nn \\
 \left(\partial_\tau -  i L\right) u^{(2)} &=& S(\tau,x; u^{(1)}) \ ,
 \eea
 where we have used again the notation in \cite{Jaramillo:2020tuu}, with $L$
 the evolution operator in Eqs. (\ref{e:L_operator_intro}) and (\ref{e:L_1-L_2_intro})
 fixed by the non-perturbed Schwarzschild spacetime. The first-order (homogeneous) equation,
 corresponds to the evolution Eq. (\ref{e:wave_eq_1storder_intro}) in the
 transient (initial data) scheme discussed in section \ref{s:pseudospectrum_close-limit}.
 In the second-order equation, the source $S(\tau,x; u^{(1)})$ would be fixed in terms
 of the first-order perturbation $u^{(1)}$. For simplicity, we write it in the following as $S(\tau,x)$.
 
 Following \cite{TreTreRed93,Schmi07} let us consider for simplicity a monochromatic
 harmonic forcing
 \bea
 \label{e:S_harmonic}
 S(\tau,x)=e^{i\omega\tau}s(x) \ ,
 \eea
  with $\omega\in \mathbb{C}$.
 We can then solve the inhomogeneous second equation in (\ref{e:close-limit_2ndorder})
 by acting with the resolvent $R_L(\omega)$ (essentially
 the Green's function of $(\omega I - L)$)  on the source $S(\tau, x)$ 
\bea
u^{(2)}(\tau,x) =\frac{1}{i}e^{i\omega\tau}\big((\omega - L)^{-1}s\big)(x) =
\frac{1}{i}e^{i\omega\tau}\big(R_L(\omega)s\big)(x) \ .
\eea
At this point, we can evaluate the maximum response $R_{\mathrm{max}}$ we can get 
in this forced system by maximizing the ratio between the norm
of $u^{(2)}$ and that of all possible sources $s$, namely
\bea
\label{e:R_max}
\!\!\!\!\!\!\!\!\!\!\!\!\!\!\!\!\!\!\!\!\!\!\!\!\!\!\!\!\!\!\!\!\!\!\!\!\!\!\!\!\!\!\!
R_{\mathrm{max}}(\omega)= \sup_{s\neq 0}\frac{||u^{(2)}||}{||s||}
= e^{-\mathrm{Im}(\omega)\tau}
\sup_{s\neq 0}\frac{||(\omega - L)^{-1}s||}{||s||} =  e^{-\mathrm{Im}(\omega)\tau}||(\omega - L)^{-1}|| \nn \\ 
\!\!\!\!\!\!\!\!\!\!\!\!\!\!\!\!\!\!\!=  e^{-\mathrm{Im}(\omega)\tau}||R_L(\omega)|| \ ,
\eea
where we have used the induced operator norm to express
the response $R_{\mathrm{max}}(\omega)$ in terms of the (induced) norm of the resolvent of $L$.
This describes a resonant phenomenon: when the frequency $\omega$ falls onto
the spectrum $\sigma(L)$ of $L$, the response $R_{\mathrm{max}}(\omega)$ diverges, characterizing
a resonance of the system. If $L$ is a normal operator,
$||R_L(\omega)||\leq 1/\mathrm{dist}(\omega, \sigma(L))$,
decreasing fast when moving away from the spectrum of $L$ and no more resonant phenomena occur.

The situation changes if $L$ is a non-normal operator. Indeed,
Eq. (\ref{e:R_max}) makes direct contact with the $\epsilon$-pseudospectrum
notion through its first characterization in (\ref{e:pseudospectrum_def}).
If $L$ is a non-normal operator then $||R_L(\omega)||$ can get very
large, therefore describing a resonant behaviour, even if $\omega$ is far from the spectrum $\sigma(L)$:
this characterizes $\omega$ as a pseudo-resonance.

The interest of the harmonic source (\ref{e:S_harmonic}) is that
any (appropriate) time-dependent source $S(\tau,x)$ can be written as a continuous
superposition of harmonic sources (\ref{e:S_harmonic}) with real $\omega$'s
through a standard Fourier transform.
Maximizing then $R_{\mathrm{max}}(\omega)$ in (\ref{e:R_max}) over all $\omega\in\mathbb{R}$
\bea
\label{e:Rmax_real}
R_{\mathrm{max}} =\sup_{\omega\in\mathbb{R}} R_{\mathrm{max}}(\omega) =
\sup_{\omega\in\mathbb{R}}||(\omega - L)^{-1}|| = \sup_{\omega\in\mathbb{R}}||R_L(\omega)|| \ ,
\eea
we get the maximum possible response to a time-dependent forcing $S(\tau,x)$. Following~\cite{TreTreRed93},
pseudospectra contour lines can then be seen as lines in the frequency-$\omega$ complex plane
with the same resonance response magnitude.
Therefore, is $\epsilon$-pseudospectra with small $\epsilon$ (therefore large $||R_L(\omega)||$)
approach the real line, pseudo-resonant phenomena can be expected.

Finally, we can assess the capability of the close-limit approximation at the second-order
in perturbation theory to account for the merger peak as a non-linear resonant
phenomenon in which second-order perturbations would act as a forcing.
This would require the pseudospectrum of Schwarzschild to present small $\epsilon$-pseudospectra
sets approaching the real line. Since this turns out not to be case (cf. 
analysis in~\cite{Jaramillo:2020tuu}) we can conclude that the second-order
close-limit approximation does neither account for the merger peak: the
close-limit approximation may apply from the merger, but not before~\footnote{Note that 
  the  Schwarzschild branch cut (continuous spectrum of $L$) does neither account
  for a resonant explanation of the merger peak, even if it takes $\epsilon$-pseudospectra
  sets to the real line, since the associated real frequency vanishes.}. This resonant
mechanism may, on the contrary, prove of interest in the dynamical
instability of ultracompact objects~\cite{BoyCarDes22}.

\bigskip

\bibliographystyle{reporthack}
\bibliography{Biblio.bib}

\end{document}